\documentstyle[11pt,aaspp4]{article}

\def\Msun{M$_\odot$}
\def\etal{{\it et al.~}}

\begin{document}

\title{ PROPER MOTION OF THE COMPACT, NONTHERMAL RADIO
SOURCE IN THE GALACTIC CENTER, SAGITTARIUS A$^*$}

\author{ D. C. Backer }
\medskip
\affil{ Astronomy Department \& Radio Astronomy Laboratory,
University of California, Berkeley, CA\\
email: dbacker@astro.berkeley.edu 
}
\bigskip
\centerline {and}
\bigskip
\author{ R. A. Sramek }
\medskip
\affil{ Very Large Array, National Radio Astronomy Observatory,
Socorro, NM\\
email: rsramek@aoc.nrao.edu}

\begin{abstract}
Proper motions and radial velocities
of luminous infrared stars in the galactic center have 
provided strong evidence for a dark mass of $2.5\times 10^6$ \Msun$~$
in the central 0.05 pc of the galaxy. The leading hypothesis
for this mass is a black hole. 
High angular resolution measurements at radio wavelengths find
a compact radio source, Sagittarius (Sgr) A$^*$, that is either the
faint glow from a small amount of material accreting onto 
the hole with low radiative efficiency or a miniature AGN core-jet system.

We provide in this paper a full report on the
first program that has measured the apparent proper motion 
of Sgr A$^*$ with respect to background extragalactic reference frame.
Our current result is:
$$\mu_{l,*}=[-6.18\pm 0.19]~~{\rm mas~y}^{-1}$$
$$\mu_{b,*}=[-0.65\pm 0.17]~~{\rm mas~y}^{-1}.$$
The observations were
obtained with the NRAO Very Large Array at 4.9 GHz over sixteen years. 
The proper motion of Sgr A$^*$ provides an estimate
of its mass based on equipartition
of kinetic energy between the hole and the surrounding stars.
The measured motion is largest in galactic
longitude. This component of the motion is consistent with the 
secular parallax that results from the rotation
of the solar system about the center, which is a global measure of
Oort's constants (A-B), with no additional peculiar motion of Sgr A$^*$.
The current uncertainty in Oort's galactic rotation
constants limits the use of this component
of the proper motion for a mass inference. In latitude we find
a small, and weakly significant, peculiar motion of Sgr A$^*$,
$-19\pm 7$ km s$^{-1}$ after correction for the 
motion of the solar system with respect to the local standard
of rest. 

We consider sources of peculiar motion of Sgr A$^*$ ranging from 
unstable radio wave propagation through intervening turbulent
plasma to the effects of asymmetric masses in the center. 
These fail to account for a significant peculiar motion.
One can appeal to an $m=1$ dynamical instability that numerical
simulations have revealed. However, the measurement of a
latitude peculiar proper motion of comparable magnitude and error 
but with opposite sign in the companion paper 
by \markcite{Reid99}Reid (\markcite{Reid99} 1999) leads us to conclude
at the present time that our errors may be underestimated, and that
the actual peculiar motion might therefore be closer to zero.

Improvement of these measurements with further observations
and resolving the differences between independent
experiments will provide the accuracies of a few km s$^{-1}$
in both coordinates that will provide both a black hole mass
estimate and a definitive determination 
of Oort's galactic rotation constants on a global galactic scale.

\end{abstract}

\section{INTRODUCTION}

The compact, nonthermal radio source, Sgr A$^*$, was discovered by \markcite{Balick74}Balick \& Brown (1974)
while looking for compact HII regions in the center of the
galaxy. The nature of Sgr A$^*$ and its role in the center of our galaxy
have been a matter of speculation over the past 25 years. 
Until recently theoretical and observational arguments were
advanced that the galactic center contains a million solar mass 
black hole that might be identified with Sgr A$^*$ \markcite{Lynden71,
Genzel94}(Lynden-Bell \& Rees 1971; Genzel, Hollenbach, \& Townes 1994). However, emission across the electromagnetic spectrum
definitively identified with, or even possibly identified with, Sgr A$^*$ 
contains no more than $10^{36}$ solar luminosities \markcite{MezgerV96, Serabyn97}(Beckert {et~al.} 1996; Serabyn {et~al.} 1997) 
which does not necessarily demand a supermassive object. 
Angular size measurements of Sgr A$^*$ also have
yet to reveal definitively the nature of this object owing to the blurring effects of
interstellar scattering in the dense, turbulent plasma near the galactic
center \markcite{Lo85, Backer88, Jauncey89, Frail94, Rogers94, Bower98, Lo98}(Lo {et~al.} 1985; Backer 1988; Jauncey {et~al.} 1989; Frail {et~al.} 1994; Rogers {et~al.} 1994; Bower \& Backer 1998; Lo {et~al.} 1998).
From the highest frequency VLBI observations we infer an upper limit 
to the size of 1 AU at
86 GHz \markcite{Rogers94}(Rogers {et~al.} 1994). Recent summaries of the variability of the radio
emission \markcite{Zhao91, Gwinn91, Backer94, Wright93, Tsuboi99}(Zhao {et~al.} 1991; Gwinn {et~al.} 1991; Backer 1994; Wright \& Backer 1993; Tsuboi, Miyazaki, \& Tsutsumi 1999) 
and limits on its linear
and circular polarization (Zhao 1992, personal communication; 
\markcite{Bower98b}Bower \& Backer \markcite{Bower98b} 1999) 
also do not give us a definitive handle
on the intrinsic nature of this object -- stellar mass object or supermassive black hole?

Over the past five years our understanding 
of both the presence of dark matter in the center and the nature of Sgr A$^*$ 
has improved radically.  Large proper motions of luminous infrared stars
within 0.1 parsec of Sgr A$^*$ have now been detected and lead
to a good estimate on the central dark mass of $2.5\times 10^6$ \Msun~
\markcite{Eckart97, Genzel97, Ghez98c}(Eckart \& Genzel 1997; Genzel {et~al.} 1997; Ghez {et~al.} 1998). However,
models for the full Sgr A$^*$ electromagnetic
flux spectrum based on low radiative 
efficiency accretion of wind-driven matter from nearby stars onto a black hole are not yet 
consistent with a mass of a few million solar masses given nominal estimates
of the mass accretion rate \markcite{Melia92a, Melia94, Falcke93, Falcke97, 
Narayan95, Narayan98, Mahadevan98}(Melia 1992, 1994; Falcke {et~al.} 1993; Falcke \& Melia 1997; Narayan, Yi, \& Mahadevan 1995; Narayan {et~al.} 1998; Mahadevan 1998). 

Shortly after the discovery of Sgr A$^*$ we began an astrometry program to 
determine its proper motion relative to extragalactic reference sources, 
active galactic nuclei and quasars, with the NRAO \footnote{The 
National Radio Astronomy Observatory is a facility of the National Science 
Foundation operated under cooperative agreement by Associated Universities, 
Inc.} Green Bank interferometer 
(\markcite{Backer82}Backer \& Sramek \markcite{Backer82} 1982; BS82).
If Sgr A$^*$ were `just' a stellar mass object, then it would be buzzing
around in the central gravitational potential well in equipartition with the other
stars and gas clouds in the center. Transverse velocity components of at
least 100-200
km s$^{-1}$ would be expected \markcite{Sellgren90}(Sellgren {et~al.} 1990).
Alternatively, if Sgr A$^*$ were indeed a 
supermassive black hole, then it might very well be at rest in the center. 
Different formation scenarios for such an object as well as considerations
of galactic dynamics predict different
residual motions of a black hole with respect to the galactic center.

Our observations from the solar system of an object in the galactic
center relative to the extragalactic sky
are sensitive, of course, to the secular parallax resulting 
mainly from the rotation of the galaxy with a small additional contribution 
from the solar motion with respect to the local standard of rest.
The expected motion is approximately 6 mas y$^{-1}$ using current values
of galactic constants \markcite{Kerr86}(Kerr \& Lynden-Bell 1986). 
If we remove this large secular parallax from the apparent motion, the
residual, peculiar motion with respect to the galactic barycenter
can be used to estimate the
mass of Sgr A$^*$  using an equipartition or other dynamical argument.
The uncertainty in the secular parallax correction is largest in
the longitude direction. Therefore,
the peculiar motion of Sgr A$^*$ in galactic latitude  is most important
for assessment of mass of the parent body of Sgr A$^*$.
Alternatively, if we assume both that Sgr A$^*$ is attached rigidly to a 
several million solar mass black hole and that this object
defines the inertial reference frame for the galaxy, then the apparent
motion can be used to define galactic constants.

The intensity distribution of Sgr A$^*$ is broadened by multipath propagation
(diffraction)
in the intervening thermal plasma whose density is perturbed on small length scales.
The detection of similar broadening of OH masers near Sgr A$^*$ by 
\markcite{Frail94}Frail {et~al.} (1994) suggests strongly that
this plasma is located in the central 140 pc of the galaxy. The apparent
diameter of Sgr A$^*$ is 1.4 mas $\lambda^{+2.0}$ where $\lambda$ is the wavelength
in cm. Past and present proper motion measurements have an error much
smaller than this size and therefore we have a particular concern about
temporal stability of diffractive and refractive propagation effects.

Our Green Bank experiment (1976-1981)
detected a proper motion in galactic longitude that was consistent
with the expected secular parallax and therefore with negligible
peculiar motion or refractive effects (BS82).
However the errors were too large to place a meaningful limit on
the mass of Sgr A$^*$. They did establish that Sgr A$^*$ was galactic and not a
chance superposition of an extragalactic background source.

In 1981 we began a new Sgr A$^*$ proper motion
experiment with the NRAO Very Large Array (VLA, see \markcite{Napier83}Napier, Thompson, \& Ekers
\markcite{Napier83} 1983).
The number of antennas, the two dimensional distribution of antennas
in a wye configuration, the excellent site for phase stability
and the sensitive receivers provided considerable new capability.
Reports of the progress of this experiment have been provided in
a series of conference reports \markcite{Backer87, Backer94, Backer96}(Backer \& Sramek 1987; Backer 1994, 1996). In this
paper we provide a full report of 8 epochs of VLA observations between
1982 and 1998. The observations are described in \S 2, and
our procedures for the determination of 
the apparent proper motion of Sgr A$^*$ are explained in \S 3. 
In \S 4 we discuss the current best estimate for galactic constants
that lead to the $\sim 6{\rm~mas~y}^{-1}$
secular parallax in galactic latitude that dominates the measured motion.
The possibility of refractive wander of the position of Sgr A$^*$
is then explored and limited by recent dual frequency data. 
The third topic in \S 4 concerns interpretation of the peculiar
motion which remains after subtraction of the estimated secular parallax.
A summary of the paper is given in \S 5.

\section{OBSERVATIONS AND DATA REDUCTION}

In 1981 we searched the literature for candidate reference sources closer
to Sgr A$^*$ than those used in the Green Bank experiment reported
previously (BS82). The Westerbork
planetary nebula searches \markcite{Wouterloot79, Isaacman81}(Wouterloot \& Dekker 1979; Isaacman 1981)
provided the most sensitive and highest angular resolution
images for location of background quasars or other compact
extragalactic sources.  We further
undertook a blind search at the VLA by making snapshot images at 5 GHz
of about 50 fields whose solid angles were determined by a combination of 
primary and delay beams. The Westerbork candidates were also observed. These 
efforts led to the identification of 3 reference sources with sufficiently
strong fluxes ($>$25 mJy) and compact structure ($<1^{\prime\prime}$).
Sources W56 (B1737-294) and W109 (B1745-291) were from the Westerbork surveys
and source GC441 (B1737-294) was the first source cataloged in our
44th blind search beam.
Figure 1 provides a map of the sky surrounding Sgr A$^*$ with the relative
locations of the three reference sources. While these sources are not
`identified' as quasars or AGNs, their brightness temperature lower limits 
and spectra are such that we can confidently assume that they were 
extragalactic. 
The three sources yield an estimated source density of 3 per 4 square degrees
at an average 5-GHz flux density of 75 mJy. This is consistent with
source counts of extragalactic sources \markcite{Condon84}(Condon 1984). A test of this
primary `extragalactic' assumption is presented in a later section. 
Table 1 provides the assumed source
positions for our primary calibration sources, Sgr A$^*$, and the three
reference sources. The initial positions and Besselian 1950 reference
frame were assumed for all measurements. In our analysis relative
offsets between the three reference sources from their assumed positions were
determined. These offsets and the improved positions for all three reference
sources are included in Table 1.

There remains a bias in the reference source positions with respect to our
primary phase calibrator, B1748-253, as will be evident when we introduce
figure 2. Furthermore our assumed position for B1748-253 at the start of
our observations is not accurate as evidence by hourly observations of
the astrometric standard B1741-038. In Table 2 we present J2000 FK5 positions
of all sources. The positions of Sgr A$^*$ and the three reference sources
were kindly provided by G. Bower \markcite{Bower99} 1999
during his polarization study. These were
referenced to B1748-253 assuming our original coordinates. The four positions
were then corrected for the errors in the original B1748-253 coordinates
as determined by that listed in the current VLA manual and that determined
in our data via the B1741-038 observations.
Our estimate of the 1$\sigma$ absolute accuracy is 5 mas. 

From 1982 to 1998 a sequence of eight observations at 4.885 GHz were conducted
using the VLA in its 36-km (A) configuration. In more recent epochs
a second band was recorded at 4.835 GHz, but this data was typically
not analyzed owing to the dominance of atmospheric errors that are very
strongly correlated between the two bands. In the
last three epochs a portion of the
standard observing schedule was devoted to observations at 8.435 GHz
and 8.485 GHz. Typically three days of observations were obtained each
epoch. In the text below we refer to the two bands by their center frequencies
of 4.9 and 8.4 GHz

Each day's observations were divided into hour-long blocks. During each block 
we first observed the 3 nearby reference sources, then Sgr A$^*$, then 2 
reference sources,
then Sgr A$^*$, then 2 reference sources, etc, with a final observation of the
3 reference sources. Every hour our phase calibration source B1748-253 and a
standard VLA calibration source B1741-038 were observed.
Table 3 gives a detailed UT schedule for a block
to show typical integration times and spacings. Identical LST stop times
were used to schedule all observations. Sgr A$^*$ was observed every 10 minutes in
these blocks.
During epochs 6 through 8 we allocated 
one LST block to 8.4-GHz observations. Given that our analysis had shown that the
dominant errors in phase referencing were temporal rather than angular we 
revised the schedule for the 8.4-GHz observations
by looking at only one reference source between
Sgr A$^*$ scans and therefore returning to Sgr A$^*$ every 6 minutes.

Table 4 provides a journal of the observations with epoch number, sequential day index
within the
epoch, calendar date, Julian Date and a code to indicate which band was observed
during which block (C for 4.9 GHz and X for 8.4 GHz). LST blocks at 15, 16, 17, 18
and 19 h were originally used with all observations at 4.9 GHz. In later epochs we observed
at 8.4 GHz during 18h block and dropped the 15h block. 
For recovery of the files containing
these observations we include in Table 4 information needed to access the archive tapes
at NRAO in Socorro, NM. 

Calibration proceeded along standard lines for the VLA. The flux densities for
3C 286 were established with the SETJY task. Then the CALIB task
was run to determine the gains for
3C 286 using recommended UV restrictions. Next CALIB was run on secondary flux
standards:
B1748-253, B1741-038, 3C 48, and NRAO 530. The flux densities of these sources were
determined using the GETJY task. The program source, Sgr A$^*$ and the 3 reference sources,
were then calibrated in flux and phase using two-point interpolation of the B1748-253
data via the CLCAL task. These calibration steps are described using the current AIPS 
program names while the earliest epochs of data were processed using predecessor
versions of the software. The hourly calibration to B1748-253 removed instrumental
phases and part of the atmospheric phase. While this helps in subsequent data
analysis, the effects of this phase  calibration are accurately removed by the 
reference source comparison described below. For epochs 1-7 this initial stage
of data reduction was done at NRAO facilities in Datil or Socorro. For the
last epoch reduction was done in Berkeley.

Our main approach to the analysis of each source observation uses the 351 
complex visibility phases for each orthogonally polarized channel
along with the time of the observation and a
table of antenna positions. These were created within the VLA DEC10 and AIPS analysis
systems by directing a matrix of scan and vector averaged phases to a disk file
using the LISTER task. This procedure
was easy to replicate each epoch as the VLA developed, and allowed us freedom
to develop algorithms for precise phase referencing. Recently we reanalysed raw
data from epochs 2 and 4 to ensure that there are no systematic errors in
these matrix listings of phases as well as the associated
times, positions and antenna locations.
We conclude that there are no systematic effects at the milliarcsecond level
in the second difference position offsets described below.

\section{ANALYSIS}
 
The phase calibration described in \S 2 using B1748-253 every hour leaves
as much as one radian of residual phase on the longer baselines with time scales for
variation as fast as 15 minutes. We reason that the bulk of this differential
atmospheric phase can be modeled by a
differential refraction angle, which is differential
owing to the B1748-253 calibration. In other words, the phases for any scan
can be modeled by a plane wave deviation from the
assumed source direction. Furthermore we expect that the differential 
refraction angles for our three reference sources will differ from those
of Sgr A$^*$ according to a simple Taylor series
expansion in angle on the sky and
in time. The discussion focuses here on atmospheric perturbations 
({\it i.e.,} tropospheric and ionospheric), 
but any source of astrometric error ({\it e.g.,} frequency, time, baseline, reference
frame) will have similar effects.
While previous experience and analysis supported this
approach, we know that the differential phase fluctuations from the
atmosphere will show higher order spatial variations than the plane-wave
assumption in this model. We expect, however, that the effects of these
higher order terms will be similarly encoded in the differential refraction
angles for the set of sources.

Our first program
reads a scan of phases and associated antenna locations, and 
then fits them to a plane wave model to produce the instantaneous refraction
angle, $\Delta {\bf s_i}(t_j)$ with an iteration algorithm. 
Sidereal time is calculated from the recorded TAI values, and current coordinates
of the sources were rotated from the B1950.0 positions (Table 1).
On the first iteration
only phase data from baselines with projected lengths between 150 and 200 k$\lambda$
are used. The minimum excludes baselines for which large scale
structure will confuse the
phase. The maximum prevents use of data that may have a $2\pi$ lobe ambiguity.
Phases that exceed 90$^\circ$ are excluded owing to a possible lobe ambiguity. On
the next iteration, the first estimate is used and the maximum baseline is 
extended to include the full array. One final pass is done to insure that the 
maximum amount of data is used. Each $\Delta {\bf s_{ij}}$ solution has its
internal error estimated from the variations of the phases. 
Much of the phase variation is not independent from baseline
to baseline, so this internal error estimate will underestimate
the uncertainty. The error will however reflect the phase
scatter, and so is useful as a relative weight in further analysis.

Figure 2 displays three sets of these differential refraction angles for
one day each in epochs 2, 3 and 8.  The data from epochs 2 and 8 show the
best differential phase stability. 
In general, the four sources, Sgr A$^*$ and the three reference objects,
meander back and forth with an amplitude of 0.1$^{\prime\prime}$ on a time scale of 
one half hour. The data from epoch 3 
(1983 September 2) display the worst differential phase
stability in the entire experiment.

One can readily see in Figure 2 that the position
of Sgr A$^*$ drifts away from the cluster of reference sources over the sixteen
year span of the data in both right ascension and declination which corresponds
to an apparent motion toward negative galactic longitude.
This drift is caused mainly by the 
inexorable rotation of the solar system around the center of the galaxy!

In our next analysis step we interpolate the differential
refraction angles of the reference source observations to the times and position of
the Sgr A$^*$ observations. Figure 2 demonstrates that 
temporal variations dominate. If we were to start the program over, we would 
choose to switch sources even more rapidly. We separately analyse the data 
within each of the one hour blocks of the schedule. Position offsets are 
removed from the reference sources as we have used constant 
source position models for all observations, 
while we determined improved positions in later years.
All reference source data in a given block, typically
12 observations, are fit to an 8th order polynomial in time.
This polynomial is then evaluated at the times of all sources
and removed. Then a simple two-point calibration is done between the data
of all reference sources for each Sgr A$^*$ observation. 
Examples of these polynomials are shown in Figure 2.

These two analysis steps can be represented as follows:
First the $\chi^2$ sums are defined that allow solution for the right
ascension ($a_k$) and declination ($d_k$) polynomial coefficients.

$$ \chi_a^2 = \Sigma_{i=1,3}[\Delta\alpha_i(t_j)-\Sigma_{k=0}^{k=n} a_k(t_j-<t>)^k]^2.\eqno(1a)$$

$$ \chi_d^2 = \Sigma_{i=1,3}[\Delta\delta_i(t_j)-\Sigma_{k=0}^{k=n} d_k(t_j-<t>)^k]^2.\eqno(1b)$$

\noindent
Then the polynomial coefficient estimates ($\hat a_k, \hat d_k$) are used to remove this
effect from all sources.

$$ \Delta\alpha_i^\prime(t_j)=\Delta\alpha_i - \Sigma_{k=0}^{k=n} \hat a_k(t_j-<t>)^k.\eqno(2a)$$

$$ \Delta\delta_i^\prime(t_j)=\Delta\delta_i - \Sigma_{k=0}^{k=n} \hat d_k(t_j-<t>)^k.\eqno(2b)$$

\noindent
Finally the primed reference source data are interpolated in time, combined
with weights to effect an interpolation in angle, 
and subtracted from the primed Sgr A$^*$ data.

$$ \Delta s_*^{\prime\prime}(t_j) = \Delta s_*^\prime(t_j) -\Sigma_{i=1}^{i=3} 
w_i~\left [ \Delta s_i^\prime(t_{j+}) \left ({t_j-t_{j-}\over t_{j+}-t_{j-}}\right )
+ \Delta s_i^\prime(t_{j-}) \left ({t_{j+}-t_j\over t_{j+}-t_{j-}}\right )\right ].
\eqno(3)$$

\noindent
The optimal weights were chosen such that the mean weighted reference position
was equal to that of Sgr A$^*$: 0.288, 0.288, and 0.424 for GC441, W56, 
and W109, respectively. One can estimate these weights by inspection
of Figure 1 which has the right ascension of W56 nearly equal to that of 
Sgr A$^*$ and
the declination of W109 nearly that of Sgr A$^*$, and the right ascension
offset of GC441 nearly double and opposite that of W109 and the declination
of GC441 nearly equal and opposite that of W56.
The errors are propagated from the internal errors carried along with the
various steps outlined above. In the best conditions these errors do indicate
the agreement of the data. The errors also display when the data is less
good, but as stated earlier, the magnitude of the errors may underestimate
the expected data agreement owing to correlation of phase errors between baselines.

Our next step is to combine the position offsets for Sgr A$^*$ for each block
on each day
using the internally propagated
errors as weighting factors. The poor quality of the low elevation
data in the 15h LST block leads us to ignore this data for all days. Only 
on a few occasions does its quality match that of the higher elevation
data. The results of this block averaging for the 4.9-GHz data
are displayed in Figure 3 along
with a weighted least squares fit for a proper motion which is:
$$\mu_{\alpha,*} = -2.70 \pm 0.15 {~~\rm mas~y}^{-1}.\eqno(4a)$$
$$\mu_{\delta,*} = -5.60 \pm 0.20 {~~\rm mas~y}^{-1}.\eqno(4b)$$
This fit is presented in Figure 4 along with the results of six other fits
to subsets of the data. In three subsets we selected one of the three
days in each epoch. In the other three subsets we selected one of the
three hour angle blocks
which are available for all epochs. These provide a measure of the effects of
the troposphere and other errors
on our measurement and serve as our primary estimator of uncertainty in the
measured proper motion.
We also explored the chance possibility that the reference sources
themselves might be galactic by setting the weight of each reference source
to 0 in separate runs. The results are all contained in the error polygon
shown in Figure 4 which is used to estimate the errors quoted above.
We conclude then that the reference sources are indeed extragalactic and not
chance compact objects in the center themselves.

The 3-epoch, 8.4-GHz data also provided a proper motion fit which is
presented in Figure 4. The result is consistent with the 4.9-GHz result
quoted above although the errors are larger. Again we used the data
on independent days to provide three independent fits to assess errors.

The phase analysis discussed here has been done primarily on computers
at Berkeley with migration from VAX to $\mu$VAX to SUN.

Use of the B1950 coordinate frame is not ideal for precision astrometry
owing to improved precession constants and its incorporation of E-terms
of aberration into the calibrator source positions. However, our differential
technique suppresses errors in the reference frame and calculation of
apparent coordinates for use in observation time modeling of the fringe
phase. We have inspected the effects of the B1950 system by precessing
source positions at and near Sgr A$^*$ from 1950 to various epochs in the range
of 1981 to 1998 with old precession and nutation values and then to 2000
using the new values as specified for FK4 to FK5 catalog conversions
by \markcite{Seidelman92}Seidelman (\markcite{Seidelman92} 1992; \S 3.5). 
We find that a false proper motion of 
$$\delta\mu_\alpha = -0.0 {~~\rm mas~y}^{-1}.\eqno(5a)$$
$$\delta\mu_\delta = -0.2 {~~\rm mas~y}^{-1}.\eqno(5b)$$
is induced. This small motion is similar for our three reference sources
and therefore has negligible effect on our measurements. The size of
the effect is significantly larger in other parts of the sky.

\section{INTERPRETATION}

\subsection{Secular parallax for object at rest in the galactic center}
The expected motion for an object at rest in the galactic barycenter, its
secular parallax, is given in galactic coordinates by 
$$[\mu_l,\mu_b]_\Pi= [\mu_l,\mu_b]_{\rm GR}+ [\mu_l,\mu_b]_\odot=
-[(A-B),0]-[V_\odot/R_\circ,W_\odot/R_\circ],\eqno(6)$$
where $A$ and $B$ are Oort's constants expressed in angular terms, $V_\odot$ and $W_\odot$ 
give the solar motion with respect to the local standard of rest in directions of
$l=90^\circ$ and $b=90^\circ$, respectively, and $R_\circ$ is the distance to the
galactic center. The 1984 IAU adopted value for $(A-B)$ is $26.4\pm 1.9$ 
km s$^{-1}$ kpc$^{-1}$ \markcite{Kerr86}(Kerr \& Lynden-Bell 1986).
More recent determinations are consistent with this value: 
\markcite{Hanson87}(Hanson 1987) uses the Lick northern sky proper motion data 
to obtain $25.2\pm 1.9$ km s$^{-1}$ kpc$^{-1}$; 
\markcite{Feast97}Feast \& Whitelock (1997) use of a Hipparcos study of cepheid stars yields $27.2\pm 1.0$ 
  km s$^{-1}$ kpc$^{-1}$; 
\markcite{Olling98}Olling \& Merrifield (1998) use a more complete model of the galactic mass field to determine
  $25.2\pm 1.9$ km s$^{-1}$ kpc$^{-1}$; 
and \markcite{Feast98}Feast, Pont, \& Whitelock (1998) analyze cepheid period-luminosity zero point from radial velocities and 
  Hipparcos proper motions and revise their previous result to $27.23\pm 0.86$ km s$^{-1}$ 
  kpc$^{-1}$. 
As the 1984 IAU value of $(A-B)$ remains in the midst of
these new estimates we will use this in further calculations. 
$$[\mu_l,\mu_b]_{\rm GR}=[-5.57\pm 0.42,0.0]~~{\rm mas~y}^{-1}.\eqno(7)$$

The solar motion has been determined by 
\markcite{Dehnen98b}Dehnen \& Binney (1998) using Hipparcos results: 
$(U_\odot,V_\odot,W_\odot)=(11.0\pm 0.4,5.3\pm 0.6,7.0\pm 0.4)$ km s$^{-1}$. 
The apparent proper motion owing
to solar motion with respect to the local standard of rest using $R_\circ =8.5$ kpc is
$$[\mu_l,\mu_b]_\odot=[-0.13\pm 0.02,-0.17\pm 0.01]~~{\rm mas~y}^{-1}. \eqno(8)$$
The total secular parallax is 
$$[\mu_l,\mu_b]_\Pi=[-5.70\pm 0.42,-0.17\pm 0.01]~~{\rm mas~y}^{-1}. \eqno(9)$$
The solar motion contributes negligible additional uncertainty to the secular
parallax.

\subsection{Apparent peculiar motion of Sgr A$^*$}

We project the observed proper motion, equation 4, from 
equatorial coordinates to galactic coordinates
and remove the expected secular parallax for an object at rest 
in the galactic barycenter, equation 9, to obtain the peculiar motion.
The north celestial pole (NCP), north galactic pole (NGP), and 
galactic center (GC) form a spherical triangle. The equatorial coordinates
of the NGP and the GC are: 12$^h$ 49$^m$ and $+$27$^\circ$ 24$^\prime$; 
and 17$^h$ 42$^m$ 24$^s$ and $-$28$^\circ$ 55$^\prime$, respectively
(B1950; \markcite{Blaauw60}Blaauw {et~al.} \markcite{Blaauw60} 1960).
The spherical angle NGP-NCP-GC is then 73.37$^\circ$ and the side of the
triangle opposite NGP-GC-NCP has length 62.60$^\circ$. NGP-GC-NCP is the
negative of the position angle 
\footnote{position angles are measured north toward east, counterclockwise
on the sky} 
of the positive galactic latitude axis ($\hat b$), 
and by law of sines is $-58.29^\circ$. The position angle of the positive
longitude axis ($\hat l$) is then +31.71$^\circ$. Errors in determination of
the galactic pole and center are of order $7^\prime$ \markcite{Blaauw60}(Blaauw {et~al.} 1960)
and hence of little consequence to these calculations. Redetermination of the
principal plane of the galaxy via population II stars seen by IRAS 
\markcite{Habing88}(Habing 1988)
would be an interesting stellar mass check on the early HI gaseous
disk determination. The resultant observed proper motion of Sgr A$^*$
in galactic coordinates is 
$$\mu_{l,*}=[-6.18\pm 0.19]~~{\rm mas~y}^{-1} \eqno(11a)$$
$$\mu_{b,*}=[-0.65\pm 0.17]~~{\rm mas~y}^{-1}\eqno(11b).$$

The observed
peculiar motion of Sgr A$^*$ is then obtained by subtracting the expected
secular parallax, equation 9, from the measurements: 
$$\Delta\mu_{l,*}=[-0.48\pm 0.46]~~{\rm mas~y}^{-1} \eqno(12a)$$
$$\Delta\mu_{b,*}=[-0.48\pm 0.17]~~{\rm mas~y}^{-1}\eqno(12b).$$
The errors have been combined in quadrature.
At a distance of 8.5 kpc the peculiar velocity of Sgr A$^*$ is 
$$v_{l,*}=[-19\pm 19]~{\rm km~s}^{-1}\eqno(13a)$$
$$v_{b,*}=[-19\pm  7]~{\rm km~s}^{-1}\eqno(13b).$$
In the subsequent sections we discuss this result further.

\subsection{Radio wave propagation effects}

VLBI observations show that the apparent angular diameter of Sgr A$^*$ depends strongly
on frequency, 1.4 mas $\lambda^{+2.0}$, which is consistent with angular 
broadening by scattering in the intervening plasma \markcite{Lo81, Lo85, Backer88, 
Jauncey89, Lo93, Alberdi93, Yusef94, Backer94, Rogers94, Bower98, Lo98}(Lo {et~al.} 1981, 1985; Backer 1988; Jauncey {et~al.} 1989; Lo {et~al.} 1993; Alberdi {et~al.} 1993; Yusef-Zadeh {et~al.} 1994; Backer 1994; Rogers {et~al.} 1994; Bower \& Backer 1998; Lo {et~al.} 1998). 
The scattering interpretation is strengthened by the demonstration that OH masers 
within 0.5 degrees of Sgr A$^*$ are similarly broadened \markcite{vanLan92, Frail94}(van Langevelde {et~al.} 1992; Frail {et~al.} 1994). 
A simple explanation is that the diffuse
thermal plasma in the central 140 pc (diameter in longitude)
is sufficiently turbulent to produce the observed
scattering. This gas may be that seen in long wavelength
thermal bremsstrahlung emission by \markcite{Mezger79}Mezger \& Pauls (1979) which
has an emission measure of at least $10^4$ cm$^{-6}$ pc. Alternatively there may be
scattering within material that is being accreted onto the black hole and serves
as fuel for Sgr A$^*$ \markcite{Backer98}(Backer \& Sramek 1999); see also \S 5 of \markcite{vanBueren78}van Bueren (1978),
a comprehensive `pre-ADAF' explanation of Sgr A$^*$ as an accreting black hole.

While most of these observations have been conducted in the northern hemisphere 
where VLBI baseline coverage is poor, several experiments 
have shown convincingly that the scatter-broadened image is elliptical with a 
ratio of axes of about 2:1 at position angle (PA) $\sim 80^\circ$. The strong 
ellipticity in scattering most likely indicates that the scattering gas is threaded 
by a relatively uniform magnetic field whose pressure dominates the thermal and 
turbulent pressures of the plasma. The thin `threads' of synchrotron emission 
detected in the galactic center provide ample evidence for strong and uniform 
magnetic fields \markcite{Yusef84}(Yusef-Zadeh, Morris, \& Chance 1984).
The field is not uniform over scales of 50 pc as the OH maser
elongations are not aligned.

Our concern here is not so much with the scattering itself, but rather with the
{\it stability} of the scattering. Consider a scattering screen located a distance $fD$ 
from Sgr A$^*$ with the observer at $D$. The screen broadens a plane wave by an 
angle $\Theta_s$. The observed source size is then $\theta_\circ=f\Theta_s$ which 
leads to a decorrelation in the visibility domain 
on baselines of length $b_\circ=1/(2\pi f\Theta_s)$. This 
decorrelation arises from phase differences through the screen on length scales
of $fb_\circ$. For 4.9 (8.4) GHz this is $240f$ $(400f)$ km. The identification of 
the scattering with a 140-pc halo around the galactic center suggests 
$f\simeq 0.01$ and therefore decorrelation on extremely small scales, 2.4 
(4.0) km. These length scales are most likely smaller than the inner scale
of the density fluctuation spectrum which is set by plasma wave dissipation
processes. When phase decorrelation occurs on scales much
smaller than that of the density fluctuations owing to many radian phase 
wrapping, the expected dependence of scattering diameter on wavelength 
is exactly $\lambda^{+2.0}$ which is consistent
with current observations. In this regime we also don't expect image wander from
large scale refractive effects, and any changes in the angular broadening
will occur on long time scales set by $\theta_\circ/v_\perp$ where $v_\perp$
is the transverse motion of the line of sight through the perturbing plasma.
We proceed to inspect the evidence for stable propagation through the
intervening medium.

The relevance of this discussion to our proper motion measurements is that our
epoch accuracy is around 1-2 mas while the scatter broadened image is
50 (18) mas (and VLA synthesized beam is 500 (180) mas) at 4.9 (8.4) GHz, 
respectively. The scattered image
size itself is very stable. 
\markcite{Lo81}Lo {et~al.} (1981) determined a size at 8.4 GHz of $17\pm 1$ mas in 1974.4
with principal resolution in the East-West direction. Later measurements
in 1983.4 had sufficient UV coverage to determine elliptical source parameters:
$15.5\pm 0.1$ mas with axial ratio of $0.55\pm 0.25$  and PA of 
$98^\circ\pm 15^\circ$ \markcite{Lo85}(Lo {et~al.} 1985). Recent VLBA observations 
provide parameters of $18.0\pm 1.5$ mas with ratio of $0.55\pm 0.14$ and
PA of $78^\circ\pm 6^\circ$
\markcite{Lo98}(Lo {et~al.} 1998). We conclude that the source size has not
changed by more than 5-10\% over 23 y either with random or secular variations.
Thus the apparent source image is not expanding or contracting at a rate 
any larger than 0.07 mas y$^{-1}$ at 8.4 GHz.

The time scale for the scattered image to sample an independent portion of the 
turbulent screen is given by the ratio of the linear
size of the image to the velocity of the screen relative to the line
of sight. If we take this transverse velocity to be 100 km s$^{-1}$, which is 
characteristic of the rotating molecular disk, then independent samples of any 
refractive beam wander (or source size change)
will occur on time intervals of 20 (7) years, respectively. 
Over the somewhat
shorter 16-y interval of our 4.9-GHz measurements, we might see just a linear 
change of the position if our above conclusion that refraction was not
important was wrong. The typical contribution 
to the proper motion from refractive wander is $0.25g$ mas y$^{-1}$, where
$g(t)$ is the fractional shift of the centroid of the scattering
disk from its long term average. Statistically the amplitude of this false 
motion would be frequency independent as the time scale shortens 
with frequency just in proportion to the apparent size. During any short time
interval the refractive motion
will differ over an octave of frequency, and so we could expect
that the effects at our two radio frequencies would differ.

A separate test of refractive effects is to look at the differential positions at
the two frequencies at a single epoch. In our Green Bank experiment (BS82)
we found that the differential positions at 2.7 and 8.1 GHz were identical to
within $\sim 0.02$ of the scattering diameter at 2.7 GHz. Note that the reference
sources in the Green Bank and VLA experiments differ. In Figure 5 we show
the differential Sgr A$^*$ positions from three days of observations in epoch 8 at
4.9 GHz and 8.4 GHz. There is a systematic offset of $\sim 5$ mas in right
ascension which is 0.1 times
the scattering diameter at 5 GHz. Source structure can be one source of difference
although 5 mas is large value for this effect.
Without further high resolution imaging and monitoring we cannot determine the 
source or the stability of this offset. A similar offset is seen in the
epoch 7 data although the errors are somewhat larger. We conclude that even if
refractive wander is present, $g$ is no more than 0.1 and the apparent motion
it might contribute is less than our current errors.

\subsection{Dynamical effects on the central black hole}
A black hole in the center of the galaxy will have a statistical motion
with respect to the galactic barycenter owing to the influence of the uneven
momentum distribution of objects surrounding it. 
Consider the motion induced by the transit, or orbit, of
a perturbing mass ($m_2$) such as a nearby star or a passing molecular cloud.
The affect of $m_2$
on the mass enclosed ($m_1(r)$), and therefore Sgr A$^*$, is given by
the acceleration $Gm_2/r^2$ acting for a time given by 
$r$ divided by the circular velocity at $r$, $r/v_c(r)$. The
circular velocity at $r$ is given by $\sqrt{Gm_1(r)/r}$. The resulting
motion of the barycenter (towards $m_2$) is then:
$$\Delta v_{\rm BC}={\sqrt{G}m_2\over\sqrt{rm_1(r)}}.\eqno(14)$$
Figure 6 shows the mass and radial distance of a number of asymmetric
masses in the center of the galaxy. In general these appear to
grow as $r^{\sim 1.5}$ which is shown in the Figure 6. The asymmetric
masses range from
the nearest solar mass star whose orbital period is long with respect
to our measurement interval to the star formation complex, Sgr B2.
Inside of about 1 pc $m_1$ is constant as shown by the IR stellar
motions \markcite{Eckart97, Ghez98c}(Eckart \& Genzel 1997; Ghez {et~al.} 1998) and $\Delta v_{\rm BC}$ will be
proportional to $r$ as one considers various contributions to the
barycentric motion. The resultant motions, however, are small, less
than 1 km s$^{-1}$. In the range of 1 pc to 100 pc the enclosed
luminous mass grows as $r^{1.2}$ based on 2-$\micron$ measurements
(see review by \markcite{Genzel94}Genzel {et~al.} (1994)). Mass asymmetries in this range then
have an influence on the barycenter motion that grows more
slowly as $r^{0.4}$. For example, the molecular cloud M-0.02-0.07
shown in Figure 6 will give the enclosed mass at its radius
a peculiar motion of about 1 km s$^{-1}$. As one goes to larger and larger
radii the peculiar motion from mass asymmetries will be increasingly
dominated by a longitude motion and not a latitude motion which
is the central concern in this paper. We conclude that 
the influence of mass asymmetries in the galactic center can
be ignored at the present level of accuracy.

The perturbations of few km s$^{-1}$ that are expected for a
central black hole in our galaxy based on the discussion above can 
be compared to that expected in other galaxies based on observed
asymmetries. The nature of the double nucleus in M31 remains uncertain.
The nucleus has probably been identified by the
large velocity dispersion at the location of the P2 nucleus
\markcite{Statler99}(Statler {et~al.} 1999). The other nucleus, P1, 
may be a concentration of stars in an eccentric disk \markcite{Tremaine95}(Tremaine 1995).
Alternatively P1
may be a star cluster which will shortly be
`absorbed' into the central by tidal disruption. 
In either case the observations indicate that the $7\times 10^7$ \Msun
black hole and the surrounding stars will not be at rest in the
mass center of M31 at the level of 10 km s$^{-1}$ owing to the
influence of the estimated $3\times 10^6$ \Msun$~$ stars in P1. This mass
asymmetry in M31 is considerably larger than that known for our galaxy
at a comparable radius (Fig. 6).

One source of the excitation of an eccentric disk in M31 mentioned
above is an unstable $m=1$ normal mode in an axisymmetric disk.
Numerical N-body simulations by \markcite{Miller92}Miller \& Smith (1992) have shown that the core
of galaxies will exhibit motions owing to an unstable $m=1$ normal mode
of oscillation. For the parameters of
our galactic center the black hole and its associated
cusp of stars could be moving as fast
as 70 km s$^{-1}$ \markcite{Miller96}(Miller 1996). The instability is the result
of an amplification of the small motions discussed above. The
direction of this putatative motion is arbitrary if the perturbations
are the result of mass asymmetries on scales less than 100 pc.
Over these scales there is as much evidence for order as disorder
with respect the well defined galactic plance seen on kpc scales.
At the level of 70 km s$^{-1}$ we definitely don't see the
effect predicted by Miller. We can be unlucky 
and the motion may be largely radial. If so, we would expect the black
hole to be offset in angle from the centroid of stars at larger radii
which could be tested with analysis of the IR stellar distribution.

If there is a massive black hole at the center of the galaxy,
\markcite{Gould98}Gould \& Ramirez (\markcite{Gould98} 1998) have shown that our limit
on the observed acceleration implies that Sgr A$^*$ is either 
coincident with or closely bound to that black hole. They point
out that acceleration has the advantage of
not being confused by uncertainty in Oort's constants. 
If one expresses both the peculiar velocity and the acceleration of 
Sgr A$^*$ in units of the Earth's motion around the Sun,
the normalized velocity and acceleration are equal at a distance
of 140 AU for a gravitational mass of $2.5\times 10^6$ M$_\odot$.
Acceleration measurements, or limits, are therefore relatively more important
for distances inside 140 AU if the measurements have comparable
precision in Earth units. If Sgr A$^*$ is a random object
in the gravitational potential that one can establish firmly from the
IR proper motion studies, then its acceleration is expected to be 
$0.27~a_\oplus$, where $a_\oplus$ is the acceleration 
of the Earth in its orbit around the sun. Our upper limit of the acceleration
allowed using the full 1982 to 1998 data set is 0.3 mas y$^{-2}$, 
or $0.06~a_\oplus$. This result, although slightly higher than that used
by \markcite{Gould98}Gould \& Ramirez is still a small compared to that of a low mass
object near the massive black hole. By comparison, the
precision of our latitude peculiar motion is 7 km s$^{-1}$, or
$~0.21~v_\oplus$. We conclude that
If the center harbors a massive black hole, then the radio source
Sgr A$^*$ must be attached to it. They also discuss the possibility
that Sgr A$^*$ is in very close orbit around the black hole with
an excursion less than our single epoch precision and orbital period
less than our time base. The VLBA result 
of \markcite{Reid99}Reid (\markcite{Reid99} 1999)
and its comparison with the longer duration VLA result here will
place further constraints on this extreme scenario. 

\markcite{Gould98}Gould \& Ramirez
also use the limit on acceleration to state the low probability
of Sgr A$^*$ being a random object passing through a dense
cluster of weakly interacting dark matter. 
In their conclusion they return to this scenario and describe
a test using flux density variations caused by Doppler boosting.
Such variations would be evident in the daily sampled data discussed
by \markcite{Backer94}Backer (1994). They note in passing that
the equipartition mass of Sgr A$^*$ based on the acceleration limit
is 250 M$_\odot$ based on a 10 M$_\odot$ characteristic mass. The
`equipartition'
mass limit for Sgr A$^*$ based on the limit on peculiar motion of $<19$ km s$^{-1}$ and 
10 M$_\odot$ IR stars moving at 1000 km s$^{-1}$ is $>2\times 10^4$ M$_\odot$.

\markcite{Maoz98}Maoz (1998) discusses the dynamical constraints on alternatives to supermassive
black holes in galactic nuclei. Critical to his discussion are estimates of the 
black hole mass and surrounding density in the cusp of stars that form around the
black hole. 
Sgr A$^*$'s diameter upper limit from the 3mm VLBI measurements of 
\markcite{Rogers94}Rogers {et~al.} (1994) is 1 AU. When combined with the mass limit this
leads to a lower limit for its density
of $\sim 10^{21}$ \Msun pc$^{-3}$. As noted by Maoz (1998 personal communication),
one can argue this point. The radio emission may come from the {\it central} 
body of a cluster or a disk and hence may not delimit the full size of the parent mass. 
In proceeding we assume that
the radio emission encompasses the parent mass as it would in the case of quasi-spherical
accretion and core-jet models. 
The density estimate is such that any
form of matter other than a black hole will have a dissipative lifetime less
than $10^8$ y.

\section{CONCLUSION}

Measurements with the NRAO Very Large Array from 1982 to 1998 at 4.8 GHz
have provided the first proper motion of the compact radio source in
our galactic center, Sgr A$^*$. The peculiar motion of Sgr A$^*$ in the
mass center of the galaxy is
obtained after removing an estimate of the secular parallax which
results from the solar motion. In latitude the estimated peculiar
motion is $19\pm7$ km s$^{-1}$.
Our ongoing uncertainty about the nature of Sgr A$^*$ leads us to use 
the limit on peculiar motion along with an equipartition argument to place a lower
bound on its mass of $2\times 10^4$  \Msun. 
The inferred mass density of Sgr A$^*$ is then $10^{21}$ \Msun pc$^{-3}$ based 
on a previous estimated 1 AU source diameter at 86 GHz. This is the highest
mass density inferred for any galactic black hole candidate. Mass density
is currently the best argument for existence of a black hole when
consideration is given to the stability of 
configurations of dark matter other than a solitary black hole.

The simplest model is that Sgr A$^*$ is radiation from the atmosphere of the
$2.5\times 10^6$ \Msun black hole. Nearly steady infall and outflow models for
the radiative properties of Sgr A$^*$ exist. The possibility of a non-zero
peculiar motion has led to consideration of the influence of known
mass asymmetries in the central region of our galaxy. We conclude that
these would account for no more than a few km s$^{-1}$ pertubation.
Another source of motion may be a $m=1$ instability in the central
potential. Our estimated peculiar motion is in fact smaller than the
estimated size of this effect although projection factors need to be
considered to make a firm statistical statement.

A nonzero proper motion
might be attributed to systematic errors in the measurements,
time variable frequency dependent effects, or variations in the
intrinsic structure. At 4.8 GHz Sgr A$^*$ is
scattered by angles significantly greater than our relative
position measurement accuracy. While one can ague that variable 
refraction is probably not important, this remains a source of 
uncertainty for the VLA measurements. 
Models for the radio emission of Sgr A$^*$ suggest an increasing 
intrinsic source size with decreasing radio frequency. This could lead
to additional systematic effects for the VLA measurements.
Further measurements at higher radio frequencies are planned to resolve these 
uncertainties. 

\acknowledgments
We commend the National Radio Astronomy Observatory and its staff for
developing and maintaining the superb Very Large Array instrument
and for providing ample support
during this extended observing campaign. The genesis of the VLA experiment
started with the 1976-1981 Green Bank experiment, and that was inspired by
a lunch time conversation with R. Fisher in Green Bank circa 1975. 
The authors support has been from UC Berkeley, NAIC, and NRAO and we therefore 
thank the NSF and California taxpayers. We thank M. Reid for discussions about
his and our measurements, and E. Maoz, A. Sternberg and I. King for comments
on the manuscript. G. Bower provided a valuable independent
analysis of absolute positions from the epoch 8 data set.

\clearpage
%% \bibliography

%% TABLES %%%%%%%%%%%%%%%%%%%%%%
\clearpage
\begin{deluxetable}{crrrr}
\footnotesize
\tablecaption{Source Positions (B1950) \label{tbl-1}}
%% \tablewidth{0pt}
\tablehead{
\colhead{Source} & \colhead{RA}   & \colhead{DEC} & \colhead{4.9-GHz Flux}
& \colhead{8.4-GHz Flux}\\
\colhead{} & \colhead{~h~~m~~s~~}   & \colhead{~~d~~m~~s~~} & \colhead{Jy}
& \colhead{Jy}
}
\startdata
B1741-038   & 17 41 20.616   & -03 48 48.90  & 2.09   & 1.75  \nl
%                                             2.46 ep2
	&	&	&	& \nl
%                                             2.50 ep4
B1748-253   & 17 48 45.792   & -25 23 17.74  & 0.507  & 0.289 \nl\nl
%                                             0.51 ep2
%                                             0.48 ep4
Sgr A$^*$     & 17 42 29.319   & -28 59 18.54  & 0.652  & 0.620 \nl\nl
%                                              .950 ep4
%                                                       .710 reid98
GC441      & 17 37 43.1110  & -29 28 20.000 & 0.035  & 0.018 \nl
%                                              .046 ep4
%                                                       .16 reid98
offset     &       -0.0107  &        +0.057 & & \nl
B1737-294  & 17 37 43.1003  & -29 28 19.943 & & \nl\nl

W56        & 17 42 42.7670  & -28 19 17.700 & 0.105  & 0.112 \nl
%                                              .066 ep4
%                                                       .36 reid98
offset     &       -0.0106  &        +0.037 & & \nl
B1742-283  & 17 42 42.7564  & -28 19 17.663 & & \nl\nl

W109       & 17 45 34.7400  & -29 06 43.400 & 0.071  & 0.072 \nl
%                                              .087 ep4
%                                                       .35 reid98
offset     &       -0.0049  &        +0.002 & & \nl
B1745-291  & 17 45 34.7351  & -29 06 43.398 & & \nl

\enddata
 
\end{deluxetable}

\begin{deluxetable}{crrrr}
\footnotesize
\tablecaption{FK5 Source Positions (J2000) \label{tbl-2}}
\tablehead{
\colhead{Source} & \colhead{RA}   & \colhead{DEC}  \\
\colhead{} & \colhead{~h~~m~~s~~}   & \colhead{~~d~~$^\prime$~~$^{\prime\prime}$~~} }
\startdata
J1751-2524 (B1748-253){\tablenotemark{a}}   & 17 51 51.2632   & -25 24 00.062  \nl
1743-0350  (B1741-038){\tablenotemark{a}}   & 17 41 58.8561   & -03 50 04.617  \nl
J1745-2900 (Sgr A$^*$){\tablenotemark{b,c}} & 17 45 40.0385   & -29 00 28.104  \nl
J1740-2929 (B1737-294){\tablenotemark{c}}   & 17 40 54.5249   & -29 29 50.290  \nl
J1745-2820 (B1742-283){\tablenotemark{c}}   & 17 45 52.4949   & -28 20 26.270  \nl
J1748-2907 (B1745-291){\tablenotemark{c}}   & 17 48 45.6841   & -29 07 39.374  \nl
\enddata
\tablenotetext{a}{From VLA CAL Manual, http://www.nrao.edu/~gtaylor/csource.html\#17}
\tablenotetext{b}{Coordinates are for proper motion epoch 1998.3}
\tablenotetext{c}{estimated 1$\sigma$ accuracy is 0.005$^{\prime\prime}$ }
\end{deluxetable}

\bigskip
\bigskip

\begin{deluxetable}{crr}
\footnotesize
\tablecaption{Observation Block Typical Schedule
\label{tbl-3}}
% \tablewidth{0pt}
\tablehead{
\colhead{Source} & \colhead{Start}   & \colhead{Stop}   
} 
\startdata
1741-038   & 09:47:00   & 09:49:20  \nl
1748-253   & 09:51:30   & 09:53:20  \nl
GC441      & 09:54:50   & 09:57:50  \nl
W56        & 09:58:20   & 19:01:20  \nl
W109       & 10:01:50   & 10:04:50  \nl
SGRACN     & 10:05:30   & 10:07:50  \nl
GC441      & 10:08:30   & 10:11:20  \nl
W56        & 10:11:50   & 10:14:50  \nl
SGRACN     & 10:15:30   & 10:17:50  \nl
W109       & 10:18:30   & 10:21:20  \nl
GC441      & 10:21:50   & 10:24:40  \nl
SGRACN     & 10:25:20   & 10:27:40  \nl
W56        & 10:28:20   & 10:31:10  \nl
W109       & 10:31:40   & 10:34:40  \nl
SGRACN     & 10:35:20   & 10:37:40  \nl
GC441      & 10:38:20   & 10:41:10  \nl
W56        & 10:41:40   & 10:44:40  \nl
W109       & 10:45:10   & 10:48:10  \nl
1748-253   & 10:48:50   & 10:52:10  \nl
1741-038   & 10:59:20   & 11:02:10  \nl
 
\enddata
 
\end{deluxetable}

\clearpage
 
\begin{deluxetable}{crrrrrrr}
\footnotesize
\tablecaption{Journal of Observations \label{tbl-4}}
\tablewidth{0pt}
\tablehead{
\colhead{Epoch} & \colhead{Day}   & \colhead{Date}   & \colhead{JD} 
  & \colhead{band} & \colhead {VLA Tape} & \colhead{Files} & \colhead{Code}
} 
\startdata
1 &1  &81mar04   &2444667.5  &CCCCC &  XL81003 & 33,34   & BACK \nl
1 &2  &81mar05   &2444668.5  &CCCCC &  XL81003 & 36,37   & BACK \nl\nl
2 &1  &82apr25   &2445084.5  &CCCCC &  XH82007 & 45,46   & BACK \nl
2 &2  &82apr26   &2445085.5  &CCCCC &  XH82007 & 48,49   & BACK \nl
2 &3  &82apr27   &2445086.5  &CCCCC &  XH82008 & 3,4     & BACK \nl\nl
3 &1  &83sep02   &2445579.5  &CCCCC &  XH83016 & 24,25,26 & AB248 \nl
3 &2  &83sep09   &2445586.5  &CCCCC &  XH83017 & 7,8     & AB248 \nl
3 &3  &83sep16   &2445593.5  &CCCCC &  XH83017 & 40,41   & AB248 \nl\nl
4 &1  &85jan05   &2446070.5  &CCCCC &  XH85001 & 18,19   & AB248 \nl
4 &2  &85jan19   &2446084.5  &CCCCC &  XH85003 & 7,8,9   & AB248 \nl
4 &3  &85jan29   &2446094.5  &CCCCC &  XH85004 & 25,26   & AB248 \nl\nl
5 &1  &86apr17   &2446537.5  &CCCCC &  XH86011 & 9,10,11 & AB388 \nl
5 &2  &86apr29   &2446549.5  &CCCCC &  XH86012 & 24,25,26 & AB388 \nl
5 &3  &86may29   &2446579.5  &CCCCC &  XH86016 & 3,4,5   & AB388 \nl\nl
6 &1  &89jan28   &2447554.5  &CCCXC &  XH89003 & 14      & AB520 \nl
6 &2  &89feb02   &2447559.5  &CCCXC &  XH89004 & 10      & AB520 \nl
6 &3  &89feb04   &2447561.5  &CCCXC &  XH89004 & 13,14   & AB520 \nl\nl
7 &1  &94apr02   &2449444.5  & CCXC &  XH94024 & 8,9     & AB708 \nl
7 &2  &94apr21   &2449463.5  & CCXC &  XH94028 & 7,8     & AB708 \nl
7 &3  &94apr26   &2449468.5  & CCXC &  XH94029 & 3       & AB708 \nl\nl
8 &1  &98apr10   &2450913.5  & CCXC &  XH98038 & 7       & AB857 \nl
8 &2  &98apr18   &2450921.5  & CCXC &  XH98040 & 9,10    & AB857 \nl
8 &3  &98apr24   &2450927.5  & CCXC &  XH98042 & 2       & AB857 \nl
 
\enddata
 
\end{deluxetable}
 
\clearpage
\setcounter{figure}{0}

\begin{figure}
\plotone{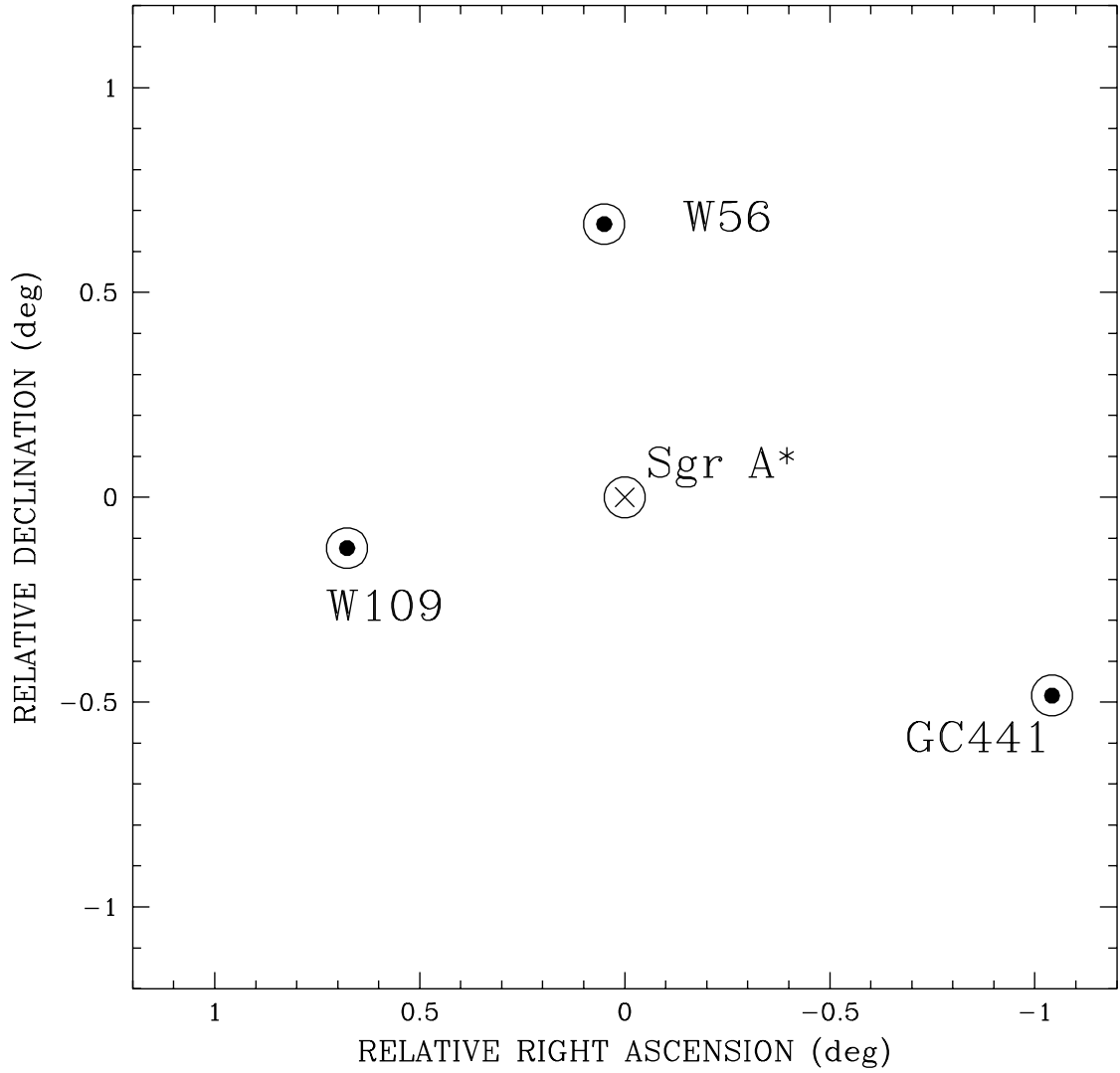}
\caption{
The relative locations of Sgr A$^*$ which is at origin and three reference
sources, W109, W56 and GC441. Observations consisted of frequent switching
between the reference sources and Sgr A$^*$.
}
\end{figure}

\begin{figure}
\plotone{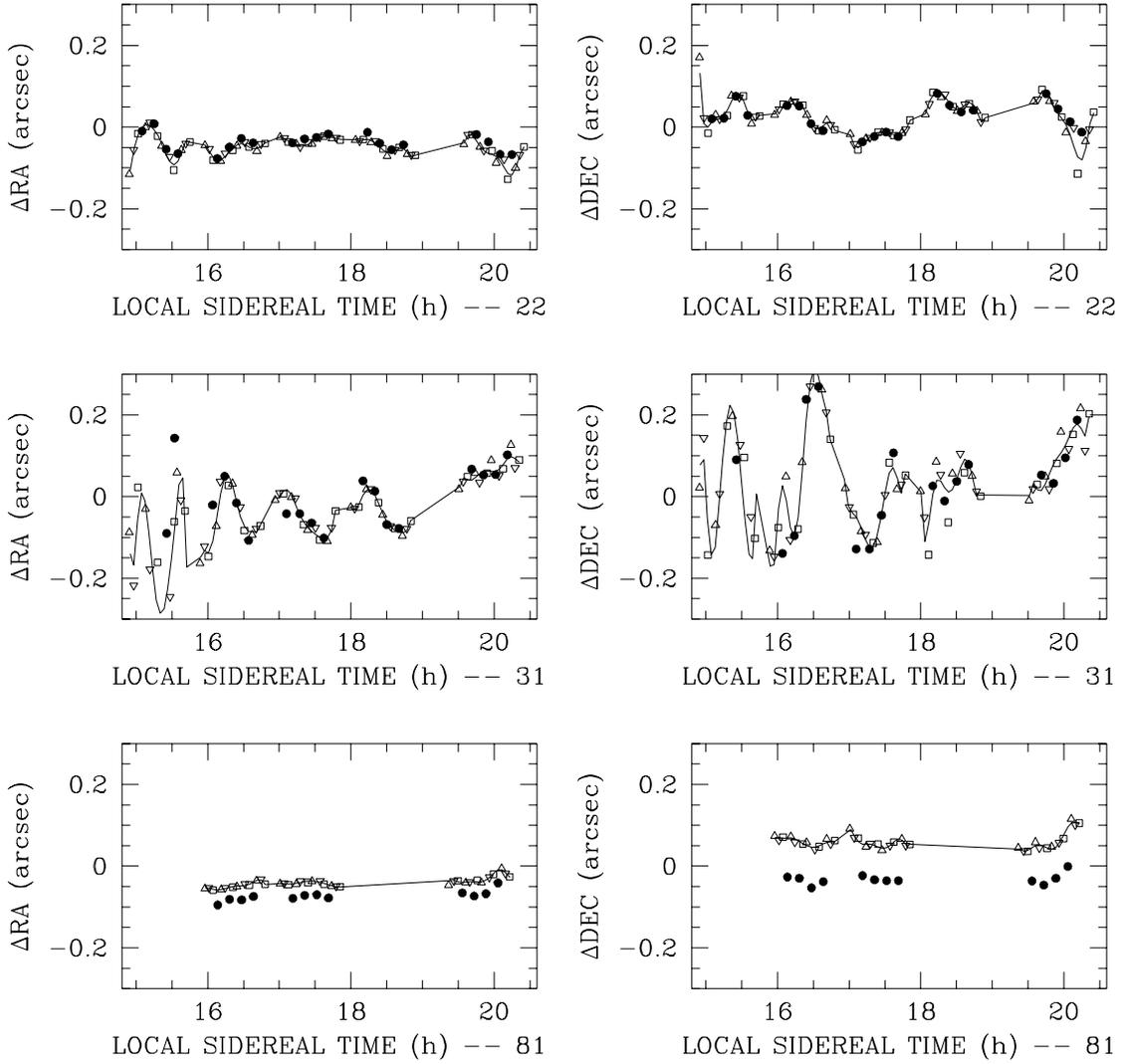}
\caption{
First difference positions (hourly phase calibration by B1748-253)
in right ascension and declination for our three
reference sources (open symbols: triangle=GC441; inverted triangle=W56;
square=W109) and for Sgr A$^*$ (solid circlular symbol) for three epochs.
The solid line represents the 8th-order temporal polynomial fit to the reference
source data. Abscissa axes provide the epoch keys: 22=epoch 2 day 2;
31=epoch 3 day 1 (a `bad' day on the Plains of St. Augustin); and
81=epoch 8 day 1 (a `spectacular' day).
}
\end{figure}
 
\begin{figure}
\plotone{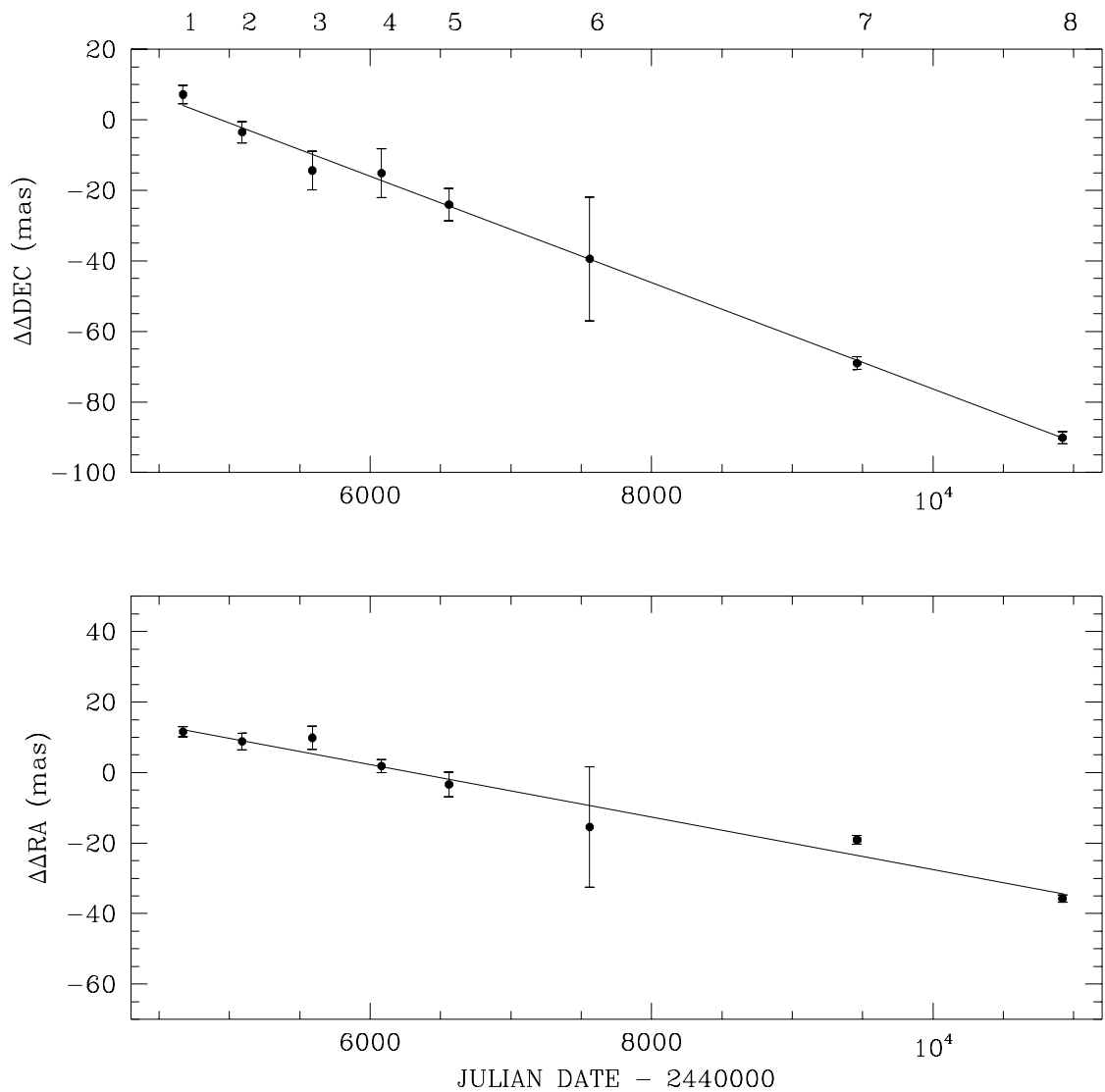}
\caption{
Weighted average second difference positions in right ascension
and declination for Sgr A$^*$ using LST blocks
at 16, 17, and 19 hours from VLA observations at 4.9 GHz. 
Errors are derived from the internal rms assuming
independence of results in different blocks and on different days. Typically
9 measurements are included. The solid line gives best fit proper motion
for Sgr A$^*$. Epochs are enumerated at top.
}
\end{figure}

\begin{figure}
\plotone{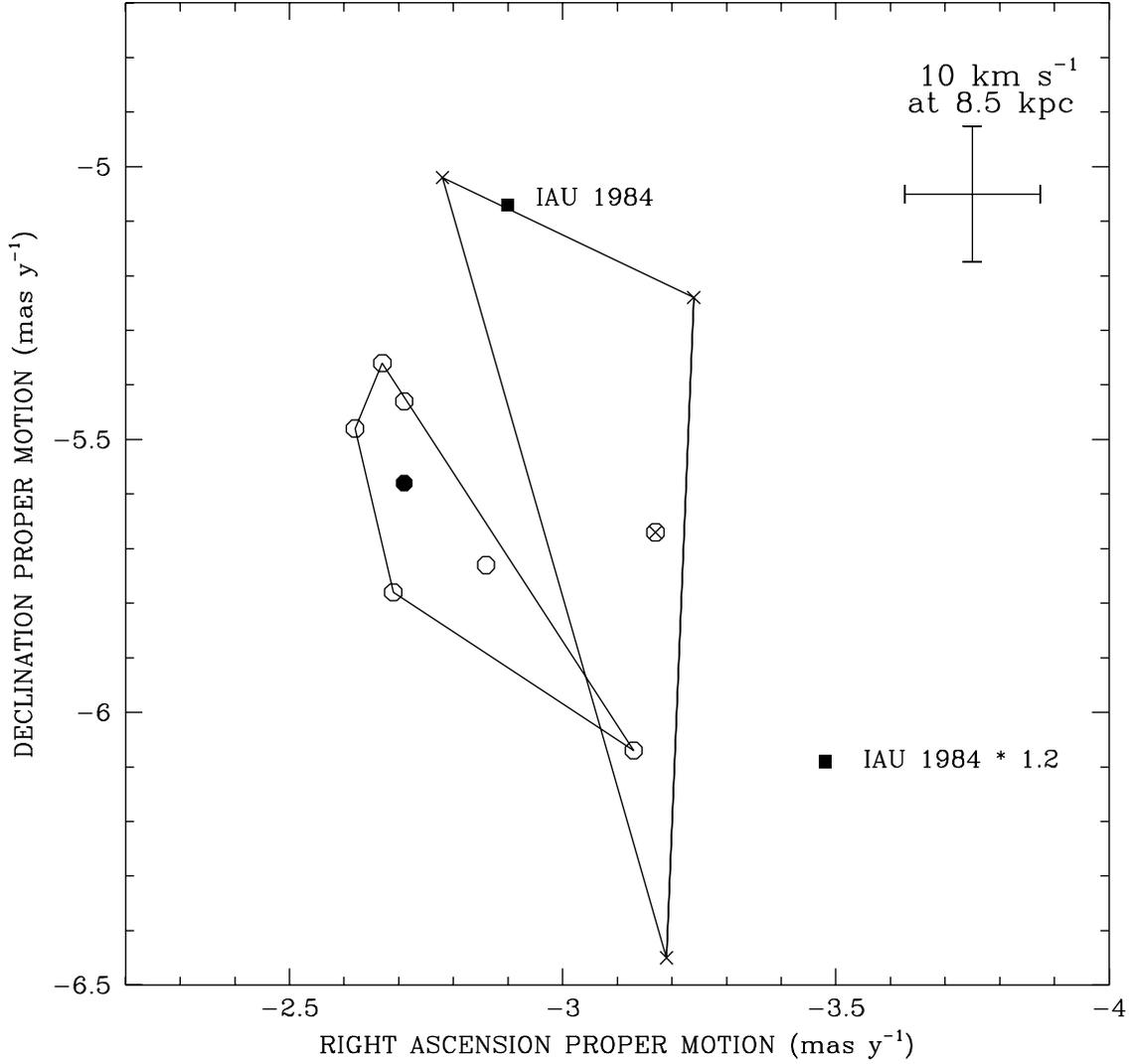}
\caption{
Proper motion estimates. Circular symbols points are based on 8 epochs (16 y)
of 5-GHz data. The solid circle is from a weighted fit to all data. 
The open circles are from subsets of the data: each of 3 hour-long blocks 
and each of three days for 5 GHz.
The $X$ symbol points are from three epochs (9 y) of 8.4-GHz data.
The circled-$X$ is from a weighted fit to all data. The plain $X$
are from each of three days. The spread
in these subset points provides our best estimate of the errors quoted in
the text for these two measurements. The solid square symbol points
give the expected
proper motion components for an object at rest in the galactic center
owing to reflex motion of galactic rotation and solar motion with
respect to local standard of rest. That labeled IAU 1984 uses
Oort's constants (A-B) in angular units given by
the 1984 IAU value (220 km s$^{-1}$ / 8.5 kpc)
summed with the solar motion. The second point labeled IAU 1984 $\times~1.2$
is the result with (A-B) increased by 20\%.
}
\end{figure}

\begin{figure}
\plotone{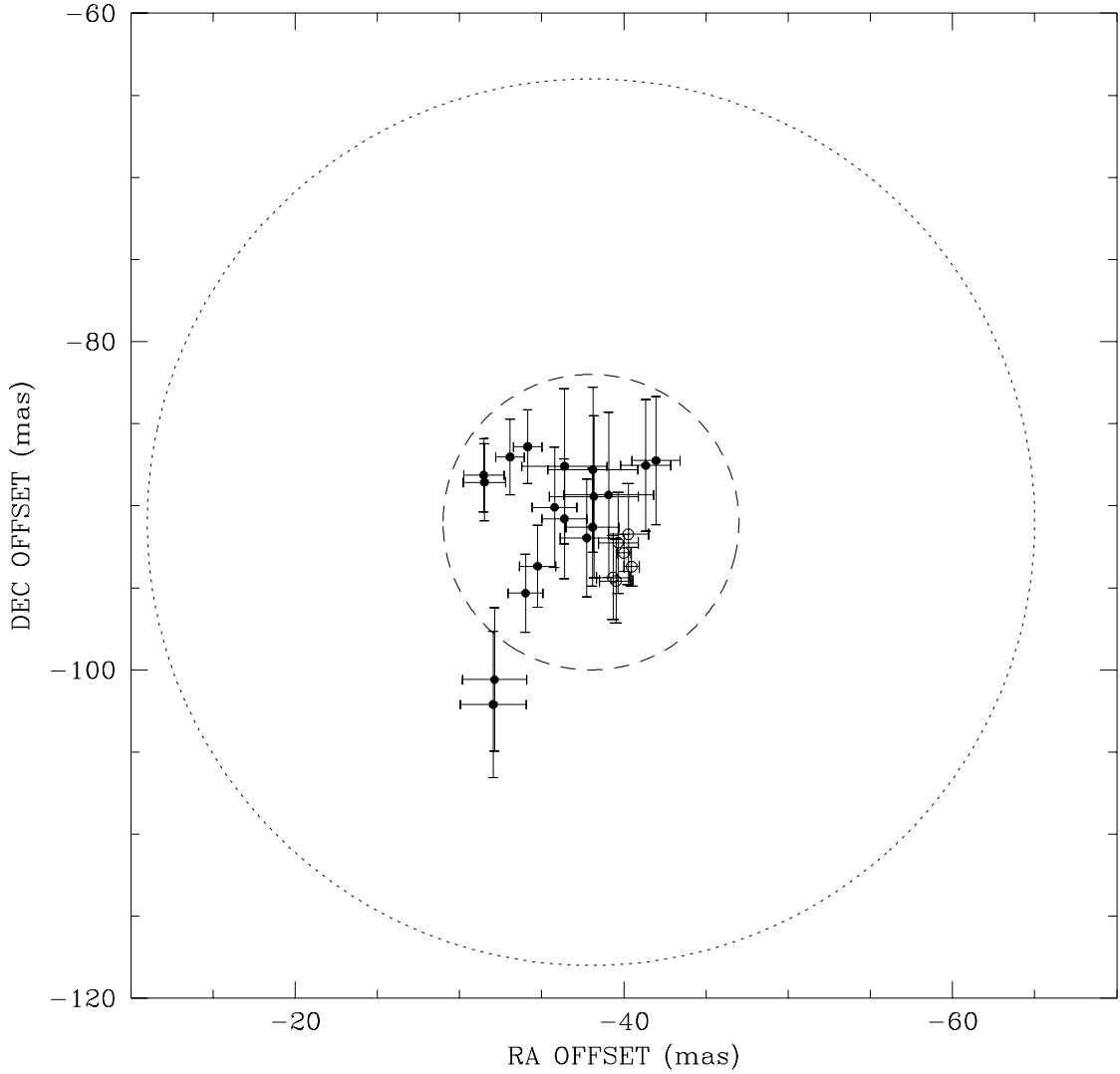}
\caption{
Second difference positions for Sgr A$^*$ from VLA observations at 4.9 GHz
(closed symbol) and 8.4 GHz (open symbol) during epoch 8 when `radio
seeing' conditions were superb. Each point represents an average over
a single 1-hour block of observations on a single day. Both VLA IFs are
processed and the very high correlation of pairs of points is obvious.
The two circles represent the diameters of the scatter broadened images
at the two frequencies.
}
\end{figure}

\begin{figure}
\plotone{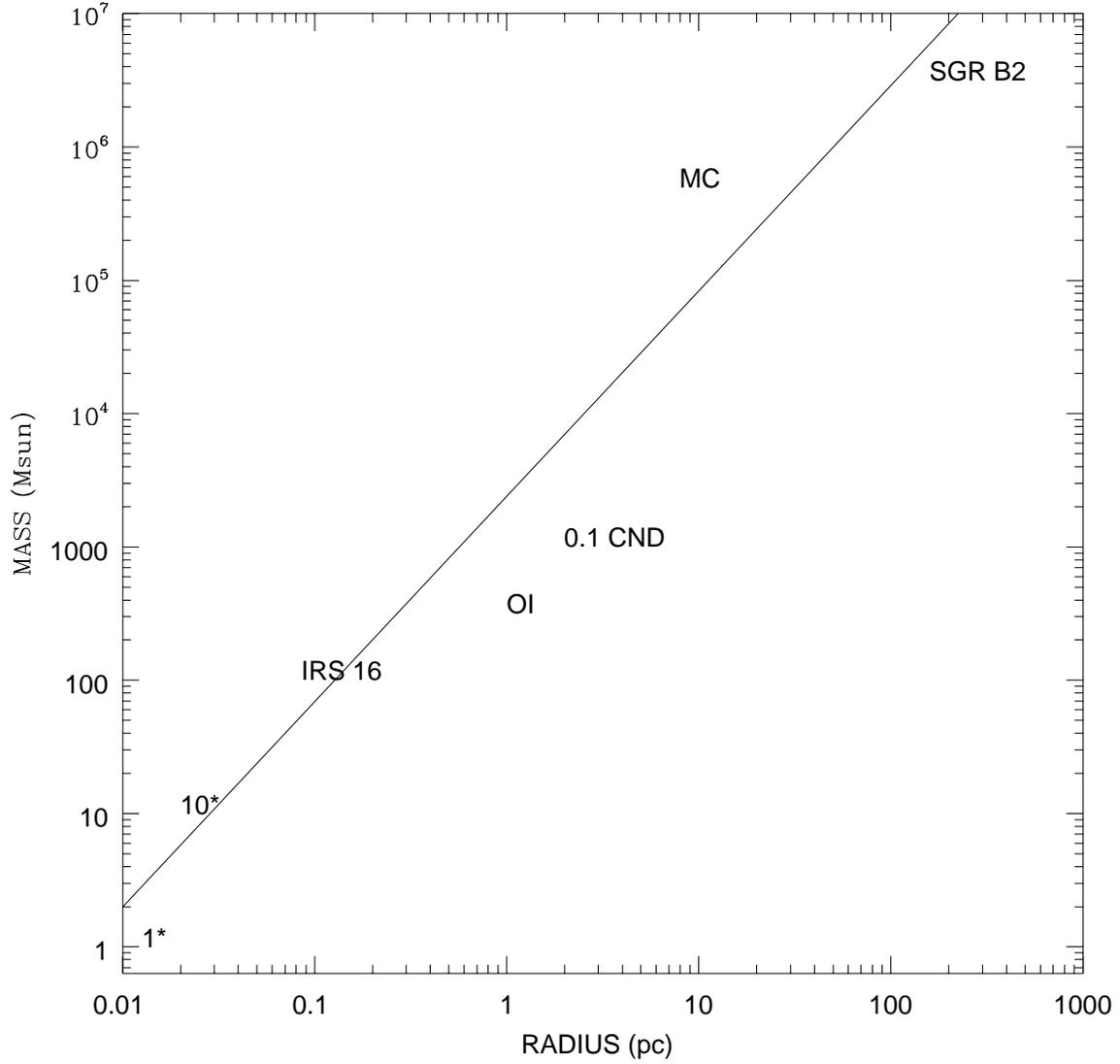}
\caption{
Asymmetric masses near the center of the galaxy as a function
of distance from Sgr A$^*$ that will affect the location and
motion of the barycenter of the galaxy. The line plotted has a slope of
1.5, $M(r)=1000r^{1.5}$ where $M$ is in solar mass units and
$r$ is in parsecs. 
}
\end{figure}


\begin{thebibliography}{}

\bibitem[Alberdi, Lara, Marcaide, Elosegui, \&  \etal 1993]{Alberdi93}
Alberdi, A., Lara, L., Marcaide, J.~M., {et al.} 1993, A\&A,  277, L1

\bibitem[Backer 1988]{Backer88}
Backer, D.~C. 1988, AIP Conf. Proc., 174, 111

\bibitem[Backer 1994]{Backer94}
---. 1994, NATO ASI, C 445, 403

\bibitem[Backer 1996]{Backer96}
Backer, D.~C. 1996, in proceedings of IAU Symposium 160 ([Reidel : Dordrecht]),  193

\bibitem[Backer \& Sramek 1982]{Backer82}
Backer, D.~C. \& Sramek, R.~A. 1982, ApJ, 260, 512

\bibitem[Backer \& Sramek 1987]{Backer87}
---. 1987, AIP CP, 155, 163

\bibitem[Backer \& Sramek 1999]{Backer98}
Backer, D.~C. \& Sramek, R.~A. 1999, in Galactic Center Workshop 98: The  Central Parsecs ([000 : 000]), in press

\bibitem[Balick \& Brown 1974]{Balick74}
Balick, B. \& Brown, R.~L. 1974, ApJ, 194, 265

\bibitem[Beckert, Duschl, Mezger, \& Zylka 1996]{MezgerV96}
Beckert, T., Duschl, W.~J., Mezger, P.~G., {et al.} 1996, A\&A, 307, 450

\bibitem[Blaauw, Gum, Pawsey, \& Westerhout 1960]{Blaauw60}
Blaauw, A., Gum, C.~S., Pawsey, J.~L., {et al.} 1960, MNRAS, 121, 123

\bibitem[Bower \& Backer 1998]{Bower98}
Bower, G.~C. \& Backer, D.~C. 1998, ApJ, 496, L97

\bibitem[Bower \& Backer 1999]{Bower98b}
Bower, G.~C. \& Backer, D.~C. 1999, in Galactic Center Workshop 98: The Central  Parsecs, ed. H.~Falcke ([000 : 000]), in press

\bibitem[Bower, Backer, Zhao, Goss, \& Falcke 1999]{Bower99}
Bower, G.~C., Backer, D.~C., Zhao, J.-H., {et al.} 1999, ApJ,  00, 00

\bibitem[Condon 1984]{Condon84}
Condon, J.~J. 1984, ApJ, 287, 461

\bibitem[Dehnen \& Binney 1998]{Dehnen98b}
Dehnen, W. \& Binney, J.~J. 1998, MNRAS, 298, 387

\bibitem[Eckart \& Genzel 1997]{Eckart97}
Eckart, A. \& Genzel, R. 1997, MNRAS, 284, 576

\bibitem[Falcke, Biermann, Duschl, \& Mezger 1993]{Falcke93}
Falcke, H., Biermann, P.~L., Duschl, W.~J., {et al.} 1993, A\&A, 270,  102

\bibitem[Falcke \& Melia 1997]{Falcke97}
Falcke, H. \& Melia, F. 1997, ApJ, 479, 740

\bibitem[Feast, Pont, \& Whitelock 1998]{Feast98}
Feast, M., Pont, F., \& Whitelock, P. 1998, MNRAS, 298, L43

\bibitem[Feast \& Whitelock 1997]{Feast97}
Feast, M. \& Whitelock, P. 1997, MNRAS, 291, 683

\bibitem[Frail, Diamond, Cordes, \& Langevelde 1994]{Frail94}
Frail, D.~A., Diamond, P.~J., Cordes, J.~M., {et al.} 1994, ApJ,  427, L43

\bibitem[Genzel, Eckart, Ott, \& Eisenhauer 1997]{Genzel97}
Genzel, R., Eckart, A., Ott, T., {et al.} 1997, MNRAS, 291, 219

\bibitem[Genzel, Hollenbach, \& Townes 1994]{Genzel94}
Genzel, R., Hollenbach, D., \& Townes, C.~H. 1994, Rep. on Prog. in Physics,  57, 417

\bibitem[Ghez, Klein, Morris, \& Becklin 1998]{Ghez98c}
Ghez, A., Klein, B.~L., Morris, M., {et al.} 1998, ApJ, 509, 678

\bibitem[Gould \& Ramirez 1998]{Gould98}
Gould, A. \& Ramirez, S.~V. 1998, ApJ, 497, 713

\bibitem[Gwinn, Danen, Middleditch, Ozernoy, \&  \etal 1991]{Gwinn91}
Gwinn, C.~R., Danen, R.~M., Middleditch, J., {et al.} 1991,  ApJ, 381, L43

\bibitem[Habing 1988]{Habing88}
Habing, H.~J. 1988, A\&A, 200, 40

\bibitem[Hanson 1987]{Hanson87}
Hanson, R.~B. 1987, AJ, 94, 409

\bibitem[Isaacman 1981]{Isaacman81}
Isaacman, R. 1981, A\&A Suppl., 43, 405

\bibitem[Jauncey, Tzioumis, Preston, Meier, \&  \etal 1989]{Jauncey89}
Jauncey, D.~L., Tzioumis, A.~K., Preston, R.~A., {et al.} 1989,  AJ, 98, 44

\bibitem[Kerr \& Lynden-Bell 1986]{Kerr86}
Kerr, F.~J. \& Lynden-Bell, D. 1986, MNRAS, 221, 1023

\bibitem[Lo, Backer, Ekers, Kellermann, Reid, \&  Moran 1985]{Lo85}
Lo, K.~Y., Backer, D.~C., Ekers, R.~D., {et al.} 1985, Nature, 315, 124

\bibitem[Lo, Backer, Kellermann, Reid, \& \etal 1993]{Lo93}
Lo, K.~Y., Backer, D.~C., Kellermann, K.~I., {et al.} 1993, Nature,  362, 38

\bibitem[Lo, Cohen, Readhead, \& Backer 1981]{Lo81}
Lo, K.~Y., Cohen, M.~H., Readhead, A. S.~C., {et al.} 1981, ApJ, 249,  504

\bibitem[Lo, Shen, Zhao, \& Ho 1998]{Lo98}
Lo, K.~Y., Shen, Z.-Q., Zhao, J.-H., {et al.} 1998, ApJ, 508, L61

\bibitem[Lynden-Bell \& Rees 1971]{Lynden71}
Lynden-Bell, D. \& Rees, M.~J. 1971, MNRAS, 152, 461

\bibitem[Mahadevan 1998]{Mahadevan98}
Mahadevan, R. 1998, Nature, 394, 651

\bibitem[Maoz 1998]{Maoz98}
Maoz, E. 1998, ApJ, 494, L181

\bibitem[Melia 1992]{Melia92a}
Melia, F. 1992, ApJ, 387, L25

\bibitem[Melia 1994]{Melia94}
---. 1994, ApJ, 426, 577

\bibitem[Mezger \& Pauls 1979]{Mezger79}
Mezger, P.~G. \& Pauls, T. 1979, in IAU Symp. 84, ed. W.~B. Burton ([Reidel :  Dordrecht]), 357

\bibitem[Miller 1996]{Miller96}
Miller, R.~H. 1996, ASP CS, 102, 327

\bibitem[Miller \& Smith 1992]{Miller92}
Miller, R.~H. \& Smith, B.~F. 1992, ApJ, 393, 508

\bibitem[Napier, Thompson, \& Ekers 1983]{Napier83}
Napier, P.~J., Thompson, A.~R., \& Ekers, R.~D. 1983, Proc. IEEE, 71, 1295

\bibitem[Narayan, Mahadevan, Grindlay, Popham, \&  \etal 1998]{Narayan98}
Narayan, R., Mahadevan, R., Grindlay, J.~E., {et al.} 1998,  ApJ, 492, 554

\bibitem[Narayan, Yi, \& Mahadevan 1995]{Narayan95}
Narayan, R., Yi, I., \& Mahadevan, R. 1995, Nature, 374, 623

\bibitem[Olling \& Merrifield 1998]{Olling98}
Olling, R.~P. \& Merrifield, M.~R. 1998, MNRAS, 295, 737

\bibitem[Reid 1999]{Reid99}
Reid, M.~J. 1999, ApJ

\bibitem[Rogers, Doeleman, Wright, Bower, Backer, \&  \etal 1994]{Rogers94}
Rogers, A. E.~E., Doeleman, S., Wright, M. C.~H., {et al.} 1994, ApJ, 434, L59

\bibitem[Seidelman 1992]{Seidelman92}
Seidelman, P.~K. 1992, Explanatory Supplement ot the Astronomical Almanac  ([University Science : Mill Valley])

\bibitem[Sellgren, McGinn, Becklin, \&  Hall 1990]{Sellgren90}
Sellgren, K., McGinn, M.~T., Becklin, E.~E., {et al.} 1990, ApJ, 359,  112

\bibitem[Serabyn, Carlstrom, Lay, Lis, \&  \etal 1997]{Serabyn97}
Serabyn, E., Carlstrom, J., Lay, O., {et al.} 1997, ApJ, 490, L77

\bibitem[Statler, King, Crane, \&  Jedrzejewski 1999]{Statler99}
Statler, T.~S., King, I.~R., Crane, P., {et al.} 1999, AJ, 00, in  press

\bibitem[Tremaine 1995]{Tremaine95}
Tremaine, S. 1995, AJ, 110, 628

\bibitem[Tsuboi, Miyazaki, \& Tsutsumi 1999]{Tsuboi99}
Tsuboi, M., Miyazaki, A., \& Tsutsumi, T. 1999, in Galactic Center Workshop 98:  The Central Parsecs ([000 : 000]), in press

\bibitem[van Bueren 1978]{vanBueren78}
van Bueren, H.~G. 1978, A\&A, 70, 707

\bibitem[van Langevelde, Frail, Cordes, \&  Diamond 1992]{vanLan92}
van Langevelde, H.~J., Frail, D.~A., Cordes, J.~M., {et al.} 1992,  ApJ, 396, 686

\bibitem[Wouterloot \& Dekker 1979]{Wouterloot79}
Wouterloot, J. G.~A. \& Dekker, E. 1979, A\&A Suppl., 36, 323

\bibitem[Wright \& Backer 1993]{Wright93}
Wright, M. C.~H. \& Backer, D.~C. 1993, ApJ, 417, 560

\bibitem[Yusef-Zadeh, Cotton, Wardle, Melia, \&  Roberts 1994]{Yusef94}
Yusef-Zadeh, F., Cotton, W., Wardle, M., {et al.} 1994,  ApJ, 434, 63

\bibitem[Yusef-Zadeh, Morris, \& Chance 1984]{Yusef84}
Yusef-Zadeh, F., Morris, M., \& Chance, D. 1984, Nature, 310, 557

\bibitem[Zhao, Goss, Lo, \& Ekers 1991]{Zhao91}
Zhao, J.-H., Goss, W.~M., Lo, K.~Y., {et al.} 1991, ASP-CS, 31, 295

\end{thebibliography}
\end{document}